\documentclass[a4paper,fleqn,usenatbib]{mnras}
\usepackage{newtxtext,newtxmath}
\usepackage[T1]{fontenc}
\usepackage{ae,aecompl}

\usepackage{graphicx}	
\usepackage{amsmath}	
\usepackage{amssymb}	
\usepackage{subfig}
\usepackage{multirow}
\usepackage{caption,booktabs}
\usepackage{deluxetable}

\newcommand{\Euclid}{{\it Euclid}}
\newcommand{\HST}{{\it HST}}

\title[New color-constrained galaxy templates for photo-z]{Improved photometric redshifts with color-constrained galaxy templates for future wide-area surveys}
\author[B. Lee et al.]{
Bomee Lee$^{1}$\thanks{E-mail:bomee@ipac.caltech.edu}
\& Ranga-Ram Chary$^{1}$
\\
$^{1}$MS314-6, Infrared Processing and Analysis Center, California Institute of Technology, Pasadena, CA 91125, USA}

\date{Accepted XXX. Received YYY; in original form ZZZ}

\begin{document}
\label{firstpage}
\pagerange{\pageref{firstpage}--\pageref{lastpage}}
\maketitle

\begin{abstract}
Cosmology and galaxy evolution studies with LSST, \Euclid, and {\it Roman}, will require accurate redshifts for the detected galaxies.
In this study, we present improved photometric redshift estimates for galaxies using a template library that populates three-color space and is constrained by \HST/CANDELS photometry. For the training sample, we use a sample of galaxies having photometric redshifts which allows us to train on a large, unbiased galaxy sample having deep, unconfused photometry at optical-to-mid infrared wavelengths. Galaxies in the training sample are assigned to cubes in three-dimensional color space, $V-H$, $I-J$, and $z-H$. We then derive the best-fit spectral energy distributions of the training sample at the fixed CANDELS median photometric redshifts to construct the new template library for each individual color cube (i.e. color-cube-based template library). We derive photometric redshifts (photo-z) of our target galaxies using our new color-cube-based template library and with photometry in only a limited set of bands, as expected for the aforementioned surveys. As a result, our method yields $\sigma_{NMAD}$ of 0.026 and an outlier fraction of 6\% 
using only photometry in the LSST and \Euclid/{\it Roman} bands. This is an improvement of $\sim$10\% on $\sigma_{NMAD}$ and a reduction in outlier fraction of $\sim$13\% compared to other techniques. In particular, we improve the photo-z precision by about 30\% at $2 < z < 3$. We also assess photo-z improvements by including $K$ or mid-infrared bands to the $ugrizYJH$ photometry. Our color-cube-based template library is a powerful tool to constrain photometric redshifts for future large surveys. 

\end{abstract}

\begin{keywords}
galaxies: distances and redshifts -- galaxies: photometry -- methods: observational, data analysis -- techniques: photometric -- surveys 
\end{keywords}

\section{introduction}

Upcoming deep, wide-area surveys with \Euclid, Nancy Grace Roman Space Telescope ({\it Roman}), and the Legacy Survey of Space and Time (LSST) of the Vera C. Rubin Observatory will cover several thousands of square degrees of sky and observe billions of sources. Robust redshift estimates of such large numbers of galaxies are required both for galaxy and cluster evolution studies and three-dimensional weak lensing analysis. Ideally, spectroscopic redshifts yield the most accurate distances.
However, spectroscopic surveys are biased towards strong emission line galaxies, are severely flux-limited, and biased against fainter galaxies. Obtaining spectroscopic redshifts of billions of objects is unfeasible, especially for faint galaxies that constitute the dominant fraction of the sample. Therefore, these future large surveys have to rely on photometric redshifts. This allows us to measure distances to most galaxies with detections in multiple bands to very faint flux density levels with good statistical accuracy. More accurate and precise photo-z estimation is a challenging task, especially for the cosmological weak gravitational lensing measurements that have a stringent redshift bias requirement ($\Delta_{z} < 0.002$; \cite{lau11}). 

Photometric redshifts are determined by developing and applying the mapping between redshift, observed flux densities and colors, using either template fitting (e.g. LePhare, EAZY, etc.) or machine-learning techniques. In recent years, many machine-learning approaches have been developed \citep{col04, car13, car14, rau15, hoy16} and  it has become popular because it is fast and powerful without the large systematic  uncertainties associated with modeling dust and emission-line contributions \citep{spe17a}. While machine-learning techniques are effective within their training sets \citep{san14, mas15, hoy15, spe17b, fot18, bil18}, the mapping between various colors (or flux densities) and redshift is not based on preexisting physical knowledge, but is newly obtained every time, using a training sample with photometry and known redshifts. Therefore, machine-learning heavily depends on a training sample with spectroscopic redshifts, which are limited to brighter sources. Besides, \cite{mas15} showed spectroscopic redshifts have thus far, missed significant regions of color space which has motivated color-space targeted redshift surveys such C3R2 \citep{mas17}.

In this study, we provide a new method to improve photometric redshift estimates using the template-fitting approach. While template-fitting approaches are fundamentally limited by the accuracy of existing templates and quality of the photometry, we implement a hybrid approach combining the methodology of both template fitting and empirical information obtained from a training sample. This relies on correlations between redshifts and observables such as colors, magnitude, and galaxy morphology. It is well-known that photometric tracers of strong features like the Lyman break at 912{\rm \AA}, Balmer break at 4000{\rm \AA}, and the $1.6 \mu m$ bump provide redshift constraints (\cite{saw02}, \cite{sal19}). It is also well-known that high redshift galaxies have fainter magnitudes and smaller sizes \cite[e.g][and references therein]{van14} than the local universe. We aim to incorporate such information about the evolution of galaxy properties in multi-dimensional parameter space obtained from a training sample, to the template-fitting approach. Instead of the usual training set with spectroscopic redshifts, our training set consists of galaxies with known photometric redshifts derived as the average of multiple fitting techniques. The accuracy of photometric redshifts depends on the quality of photometry (high spatial resolution and high signal-to-noise) and wide spectral coverage with many photometric bands sampling the spectral energy distribution \citep[SED;][]{bra08, dah10}. For our purpose, the \HST/Cosmic Assembly Near-infrared Extragalactic Legacy Survey \citep[CANDELS;][]{gro11, koe11} is the optimal data set to train galaxy SED templates as it has excellent spectral coverage from the X-rays to far-infrared with exceptional depth and spatial resolution. \cite{fot18} presented a similar hybrid approach identifying the optimal photometric redshift model library for each source class before performing SED fitting, but they use a machine-learning algorithm trained with spectroscopic redshifts to classify the model templates. They found that their method is advantageous for reducing catastrophic outliers for all types of galaxies and deriving photometric redshifts for AGN and QSOs, while the photometric redshift scatter ($\sigma=0.033-0.038$) is slightly worse compared to the pure template fitting and machine-learning methods.

Here, we do not aim to invent new template models or a fitting code. We leverage \HST/CANDELS photometry to constrain the range of templates and apply those to derive improved photometric redshifts for future wide area surveys. The structure of this paper as follows. CANDELS data is introduced in Section 2 together with selections of the training and target sample. We describe the method used to construct the new template library which populates three-dimensional color space, and derive the photometric redshifts using our new template library in Section 3. In Section 4, we apply our method to the photometry expected from the LSST and \Euclid-deep/{\it Roman} fields and validate the photometric redshift performance. We conclude in Section 5. 

All magnitudes are expressed in the AB system unless stated otherwise. We use a standard $\Lambda$CDM cosmology with $H_{0}$=70 km s$^{-1}$ Mpc$^{-1}$, $\Omega_{\rm M} = 0.3$, and $\Omega_\Lambda = 0.7$, which is broadly consistent with the recent results from \textit{Planck}.
\section{CANDELS Data}

We use \HST/CANDELS \citep{gro11, koe11} observations that span five well-studied extragalactic fields: 
the Great Observatories Origins Deep Survey (GOODS) Northern and Southern
fields (GOODS-North and GOODS-South), the UKIRT InfraRed Deep Sky Surveys (UKIDSS) Ultra Deep Field (UDS), the Extended Groth Strip (EGS) field,  the Cosmic Evolution Survey (COSMOS) field. 
These fields together map out about 900 arcmin$^2$ of sky achieving 5 $\sigma$ point source limiting depth of $\sim27.3$ AB mag (wide) and $\sim 28.0$ AB mag (deep) in the H (F160W) band.
CANDELS multi-wavelength photometry catalogs span from the U-band through 8 $\mu m$ and are publicly available in each field (GOODS-S; \cite{guo13}, GOODS-N; \cite{bar19}, UDS; \cite{gal13}, EGS; \cite{ste17}, COSMOS; \cite{nay17}). Each field of CANDELS has been observed with many different telescopes and instruments in many different wavelengths. The multiple bands we use in this study for each field are listed below.
\begin{itemize}
\item GOODS-S (GS): 17 bands with U(Blanco/Mosaic II or VLT/VIMOS), F435W, F606W, F775W, F814W, F850LP from \HST/ACS, F098M, F105W, F125W, F160W from \HST/WFC3, Ks (VLT/ ISAAC and HAWK-I), 3.6$\mu m$, 4.5$\mu m$, 5.8$\mu m$, 8.0$\mu m$ from Spitzer/IRAC. 
\item GOODS-N (GN): 16 bands with U(KPNO/Mosaic), F435W,  F606W, F775W, F814W, F850LP from \HST/ACS, F105W, F125W, F140W, F160W from \HST/WFC3, Ks (Subaru/MOIRCS), K (CFHT/MegaCam), 3.6$\mu m$, 4.5$\mu m$, 5.8$\mu m$, 8.0$\mu m$ from Spitzer/IRAC. 
\item UDS: 17 bands with U (CFHT/MegaCam), B, V, R, i, z from Subaru/Suprime-Cam, F606W, F814W from \HST/ACS, F125W, F160W from \HST/WFC3, Y and Ks from VLT/HAWK-I, K from UKIRT/WFCAM, 3.6$\mu m$, 4.5$\mu m$, 5.8$\mu m$, 8.0$\mu m$ from Spitzer/IRAC. 
\item EGS: 20 bands with u, g, r, i, z, from CFHT/ MegaCam, F606w, F814W from \HST/ACS, F125W, F140W, F160W from \HST/WFC3, J1, J2, J3, H1, H2 from Mayall/NEWFIRM, Ks from CFHT/WIRCAM, 3.6$\mu m$, 4.5$\mu m$, 5.8$\mu m$, 8.0$\mu m$ from Spitzer/IRAC. 
\item COSMOS: 33 bands with u, g, r, i, z from CFHT/MegaPrime, B from Subaru/Suprime-Cam, F606W, F814W from \HST/ACS, F125W, F160W from \HST/WFC3, Y and  Ks from VISTA/VIRCAM, J1, J2, J3, H1, H2 from Mayall/NEWFIRM, 3.6$\mu m$, 4.5$\mu m$, 5.8$\mu m$, 8.0$\mu m$ from Spitzer/IRAC, and 12 intermediate bands  in optical (IA484-IB827) from Subaru/Suprime-Cam. 
\end{itemize}

We correct the photometry in the catalogs
for Galactic extinction using the value given by the IRSA Galactic dust reddening and extinction calculator\footnote{https://irsa.ipac.caltech.edu/applications/DUST/}. Extinction values are calculated at the center of each field based on \cite{sch11} and are provided for a small set of filters. We then interpolate between the values to determine the extinction at the central wavelength of each filter in our data set. The median g-band extinction in the 5 fields is 0.040, 0.026, 0.061, 0.074, 0.027 mag for GN, GS, COSMOS, UDS and EGS, respectively.
We note that applying the Galactic extinction correction slightly reduces the photometric redshift scatter by about $2-3\%$ within the CANDELS fields. 

\begin{figure}
\centering
\includegraphics[width=\columnwidth]{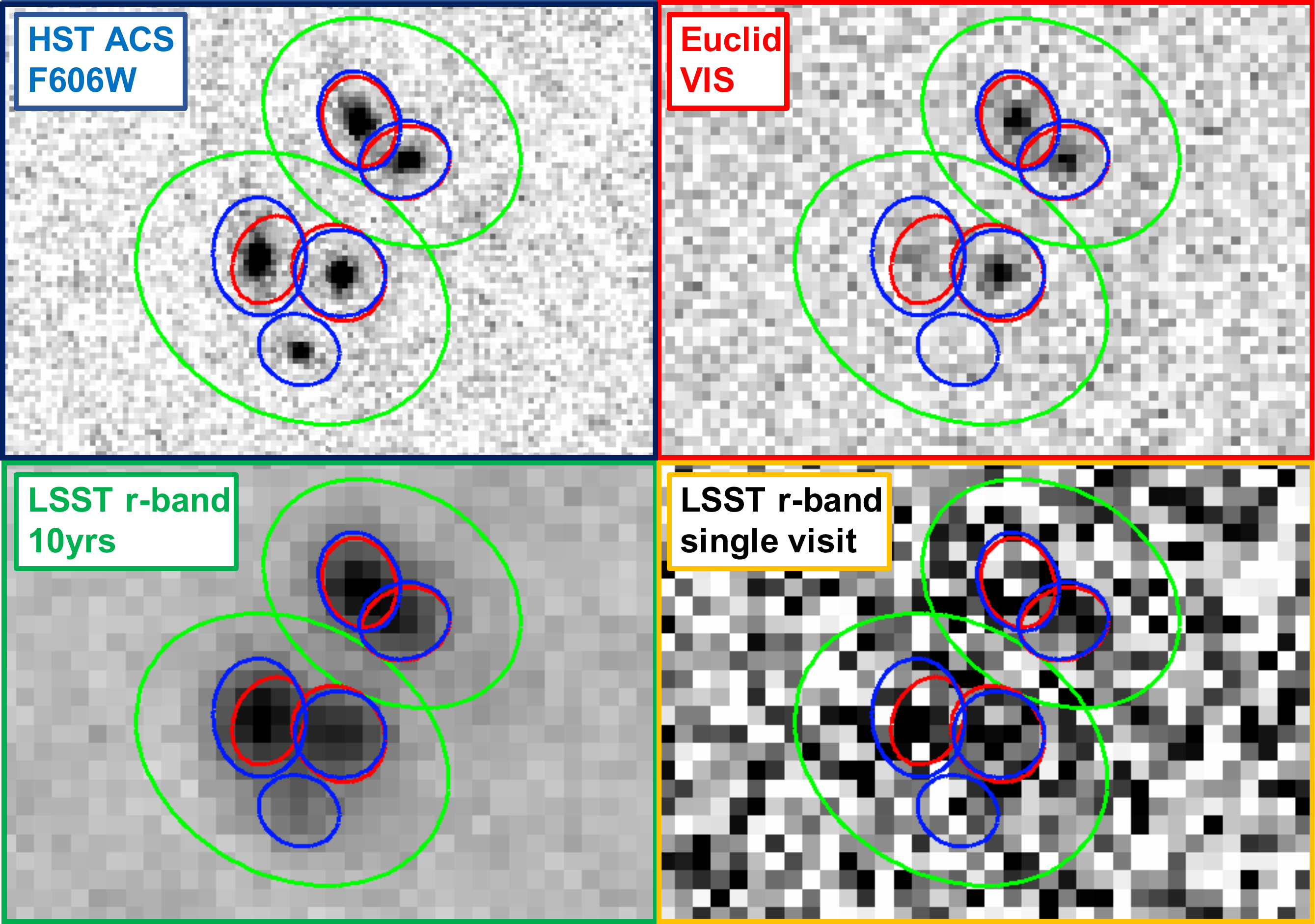}\vspace{-0.2cm}
\caption{Source confusion in an optical band (centered at $\sim$6000{\rm \AA}), along with the isophotes derived from photometry on each of the images. The green isophotes are derived from ground-based resolution r-band data of 27.7 AB mag (e.g. LSST), the red isophotes are from the {\it Euclid}-quality optical data while the blue isophotes are from the \HST/ACS data. The sources are barely detected in the LSST single epoch data which has
a depth of 24.7 AB mag. In the absence of the deeper, space-resolution data, source confusion would result in both erroneous shape and photometry estimates in ground-based data and also affect catalog matching. }
\label{fig:blended}
\end{figure}

It is worth noting that at depths beyond $\sim25$\,AB mag in the optical, source confusion in seeing-limited imaging biases photometry \citep[See Figure~\ref{fig:blended}; e.g.][]{JSPreport}. The {\it Hubble}-quality spatial resolution of the deep optical data in the catalogs, combined with the priors on morphology and galaxy shapes applied for cataloging 
alleviates much of the confusion \citep{mer16}. Thus, by using the CANDELS catalogs, we have the highest quality photometry available to build up the spectral energy distribution of the training sample of galaxies.

\subsection{Training Sample}

The galaxies in the training sample are required to have signal-to-noise (S/N) $> 3$ in U, B (F435W), F606W, F814W, F125W, F160W, K (or Ks), and 3.6$\mu m$ IRAC 1 channel to avoid having photometric uncertainties dominate redshift estimates. We then exclude suspicious sources and stars based on flags provided from the CANDELS photometry catalog; 
\begin{itemize}
\item Suspicious sources: SExtractor PhotFlag $=0$
\item Stars: CLASS\_STAR from SExtractor $\ge 0.98$
\end{itemize}
SExtractor PhotFlag is used to designate suspicious sources that fall in regions of the image which are contaminated by stellar spikes or halos or edge effects \citep{guo13}. By using PhotFlag=0, we are able to exclude detections of star spikes, halos, and bright stars, as well as excluding sources that are either artifacts or falling at the edge of the image.
We also exclude AGN-dominated sources identified from X-ray, IR, and/or radio detections (using the CANDELS catalog \textrm{AGN\_Flag}) 
which are about 1\% of the sample. Since we do not have AGN templates to constrain their redshifts correctly, those sources usually result in higher uncertainties in the photometric redshifts compared to more normal galaxies \citep{sal09}. 

To build as large a training set as possible, which is not biased towards strong emission line sources,
we consider only galaxies having photometric redshifts, i.e. we exclude all objects with spectroscopic redshifts from the training sample. For that, we use the CANDELS photometric redshifts (photo-z), which is the median redshift from 11 different photometric redshift codes, each using different template spectral energy distributions, and priors. \cite{dah13} showed that the median redshift results in the best estimates among different photo-z codes. This leaves us with a total of 39,391 galaxies as the training sample from all five CANDELS fields. 

\begin{figure}
\centering
\includegraphics[width=\columnwidth]{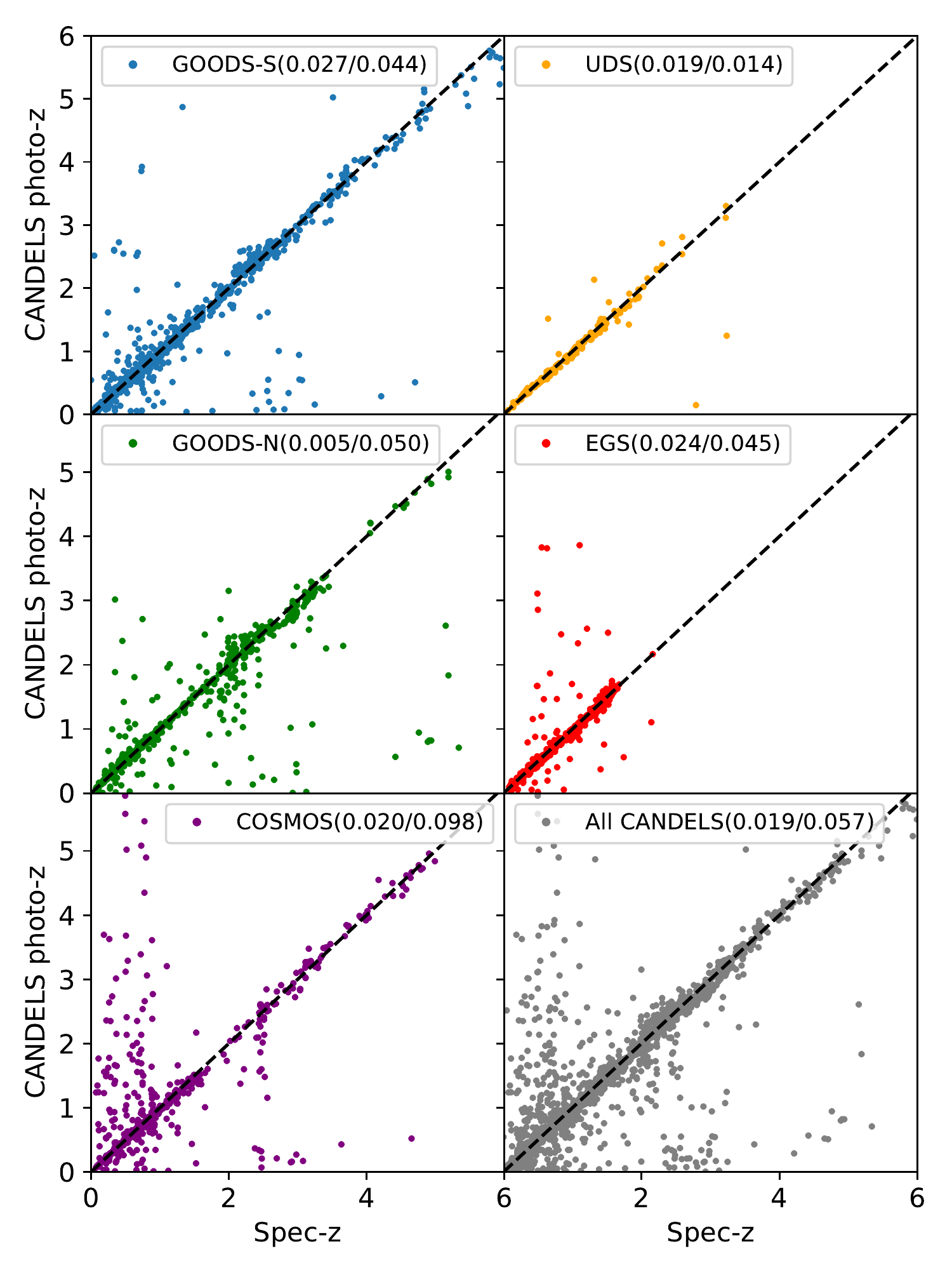}\vspace{-0.5cm}
\caption{CANDELS median photometric redshifts (photo-z) vs. spectroscopic redshifts (spec-z) in five CANDELS fields. Galaxies are color-coded by fields; GOODS-S (blue), GOODS-N (green), UDS (orange), EGS (red), COSMOS (purple), and All fields (grey). The numbers in parenthesis are $\sigma_{NMAD}$ and outlier fraction of each field. There is a relatively low fraction of high redshift galaxies beyond $z_{spec}=2$, which is about 12\% the sample with spec-z, and most of them (80\%) are in the GOODS-S and -N fields. } 
\label{fig:candels_photoz}

\end{figure}

Figure \ref{fig:candels_photoz} shows comparisons between the spectroscopic redshifts and CANDELS median photo-z in five fields with different colors. Spectroscopic redshifts are provided by CANDELS and are selected to have a spec-z quality flag $=1$ which means the spectroscopic data is secure. In addition to 381 spec-zs obtained from the CANDELS COSMOS catalog, we have collected 972 publicly available spec-zs in the CANDELS-COSMOS field from zCOSMOS \citep{lil07}, LEGA-C \citep{van16}, VUDS \citep{tas17}, COSMOS DEIMOS \citep{has18}, and C3R2 \citep{mas19} surveys. In total, we obtain 5,539 galaxies in five CANDELS fields after excluding suspicious sources, stars, and AGN candidates using the same flags mentioned in Section 2.1. Including all five fields, the CANDELS median photo-z has yielded a photo-z precision of $\sigma_{NMAD} = 0.019$ and a catastrophic outlier of 5.7\% at $0 < z < 6$ (see the definitions of statistics in Equation 1). $\sigma_{NMAD}$ of GOODS-N are very small with 0.005 because of an inclusion of medium-width optical bands \citep{bar19}. It is worth noting that the photometric redshift accuracy reported here may not represent galaxies in the entire training sample because the photo-z precision and outlier fractions are calculated using a subsample of galaxies with spectroscopic redshifts that are brighter in most case and at lower redshift compared to the full galaxy sample. The accuracy of photometric redshifts are expected to degrade with fainter magnitudes \citep{ilb09, dah13}.

There are pros and cons of using a photo-z selected galaxy sample for developing the training set. 
Spectroscopically confirmed galaxies tend to be fewer in number, and biased towards bright, emission line galaxies.
This can be problematic, especially, for future large photometric surveys that will image billions of galaxies including a large fraction of faint galaxies.
Photo-z selected galaxies are an order of magnitude larger in number density, have less redshift accuracy but are relatively unbiased towards galaxies of a particular type.
This alleviates the impact of spectroscopic misclassification, which tends to rely on single line redshifts. 
However, we are not immune from any systematic uncertainties inherent in CANDELS median photo-zs that introduce an additional scatter to our photo-z estimates. 

\begin{figure}
\centering
\subfloat{
\includegraphics[clip, width=0.8\columnwidth]{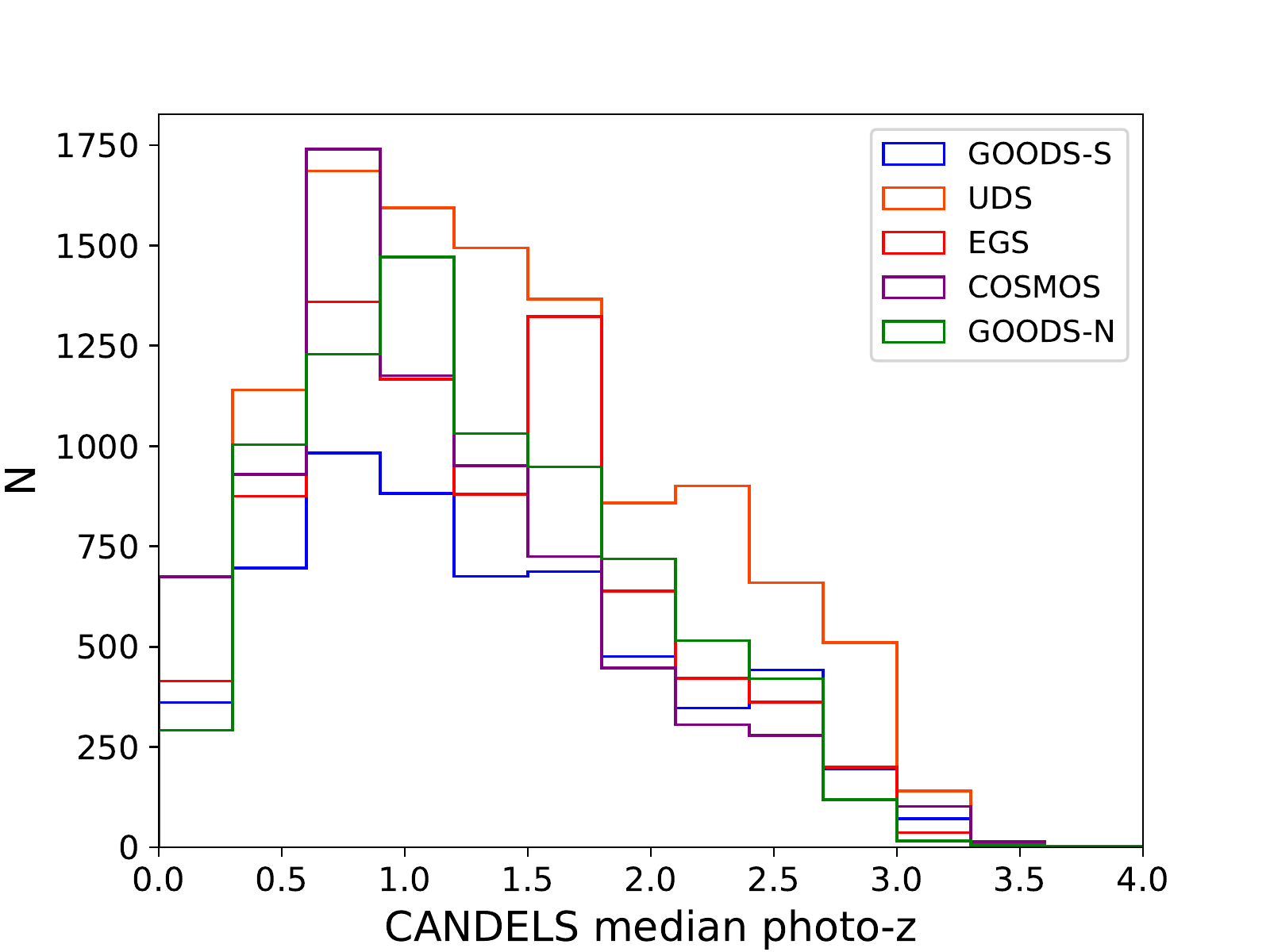}
}
\vspace{-0.3cm}
\subfloat{
\includegraphics[clip, width=0.8\columnwidth]{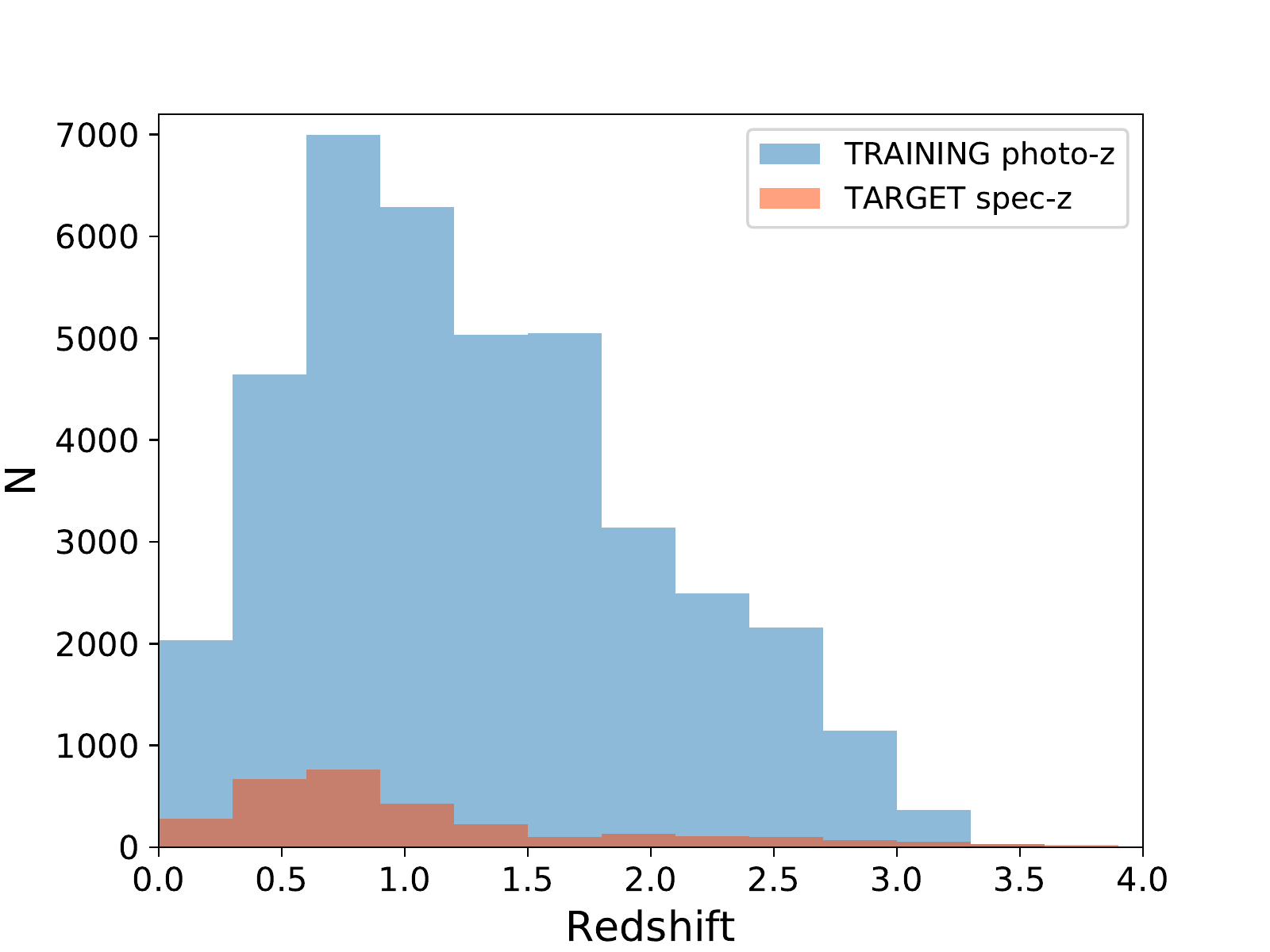}
}
\caption{Top: Photo-z distributions of the training sample in five fields, GOODS-S (blue), GOODS-N (green), UDS (orange), EGS (red), COSMOS (purple). Bin size of the histogram is $\Delta z =0.3$. The constraint of S/N$ > 3$ in the U-band, restricts galaxies in the training sample to $z < 3.6$. Note that the red end of CTIO U-band in GOODS-S extends to 4120$\textrm{\AA}$. Bottom: The photometric redshift distribution of the training sample with a blue histogram and the spectroscopic redshift distribution of the target sample with an orange histogram. Target sample is chosen with galaxies having spectroscopic redshifts at $0 < z < 6$ in the GOODS-S and - N fields and having ugrizYJH filters close to LSST-10 years+\Euclid-deep/{\it Roman}-quality photometry.}
\label{fig:zdist}
\end{figure}

In Figure~\ref{fig:zdist}, we show the distribution of CANDELS median photo-zs of galaxies in the training sample. In the top panel, the training sample in different fields have photo-z $< 3.6$ because we require 3 sigma detection in U-band. The red end of the U-band filter in CANDELS extends to $\sim 4120\textrm{\AA}$, corresponding to the redshifted Lyman-break at $z\sim 3.6$. 

\subsection{Target Sample}

\begin{table*}
\centering
\caption{Comparison between \HST/CANDELS and Future surveys with LSST, \Euclid, and {\it Roman}. \label{tab:survey}}
\begin{tabular}{cccc}
\hline \hline
Project & Wavelength range & 5$\sigma$ sensitivity [AB mag] & Spatial Resolution [arcsec] \\
\hline
\HST/CANDELS & VI(ACS) YJH (WFC3) & $> 27.4$ & 0.1(VI)/ 0.18(YJH)  \\ \hline
LSST\textsuperscript{a} & $ugrizy$ & $< 24.7$(single)/$< 27.7$(10 years) & 0.7 \\  \hline
\Euclid NIR\textsuperscript{b}& $YJH$ &  24(wide)/  26(deep) & 0.3  \\ \hline
{\it Roman}\textsuperscript{c} & $zYJH$+F184 &  $< 26.7$ & 0.18 \\ \hline
\multicolumn{4}{l}{\textsuperscript{a}\footnotesize{https://docushare.lsst.org/docushare/dsweb/Get/LPM-17}} \\
\multicolumn{4}{l}{\textsuperscript{b}\footnotesize{Euclid red book \citep{lau11}}} \\
\multicolumn{4}{l}{\textsuperscript{c}\footnotesize{WFIRST AFTA 2015 Report \citep{spe15}}}
\end{tabular}
\end{table*}

Future large extragalactic surveys will observe several billions of galaxies and provide multi-wavelength catalogs consisting of optical ($ugrizy$ from LSST) and near-infrared ($YJH$ from {\it Euclid}\footnote{{\it Euclid} will also image in a broad optical band (VIS) from 550$-$900nm.} , with the addition of F184 from {\it Roman}) photometry (See Table~\ref{tab:survey}). The precision of photometric redshifts derived using only these bands will be limited, since it will be challenging for ancillary data sets (e.g. narrow bands) to achieve the unprecedented sensitivities 
and areal coverage of these surveys. Since our primary goal is to assess the quality of redshifts that can be derived from these surveys, we select 1,696 galaxies in the GOODS-S and 1,355 galaxies in the GOODS-N fields with spectroscopic redshifts at $0 < z < 6$ as our target sample as shown in the bottom plot of Figure~\ref{fig:zdist}. We only use photometry in 8 filters, including U from the ground-based telescopes, F435W, F606W, F814W, F850p from \HST/ACS, and F105W, F125W, F160W from \HST/WFC3, to derive the photo-zs in Section 4.1. We also assess the quality of photo-zs of the target sample with an addition of the IRAC 1 channel in Section 4.2. 

The photometry used here can be considered as a case study for LSST-10 years+\Euclid-deep/{\it Roman}-quality photometry albeit they are about a magnitude deeper, except U-band as shown in Table~\ref{tab:survey}. We note that the spatial resolution of the data used here is far superior to that arising from LSST (0.1\arcsec ~ vs. 0.7\arcsec~FWHM), minimizing the impact of source confusion on photometric bias and thereby redshift errors.

In future work, we will apply the constraints from our training set to a completely independent spectroscopic redshift sample, such as that derived from the C3R2 survey \citep{mas17}; however those fields do not have consistently photometered data and deep near-infrared imaging such as that available in the CANDELS field.

\section{Method: Reconstructing galaxy templates in color space}

\begin{figure*}
\centering
\includegraphics[width=5.5in]{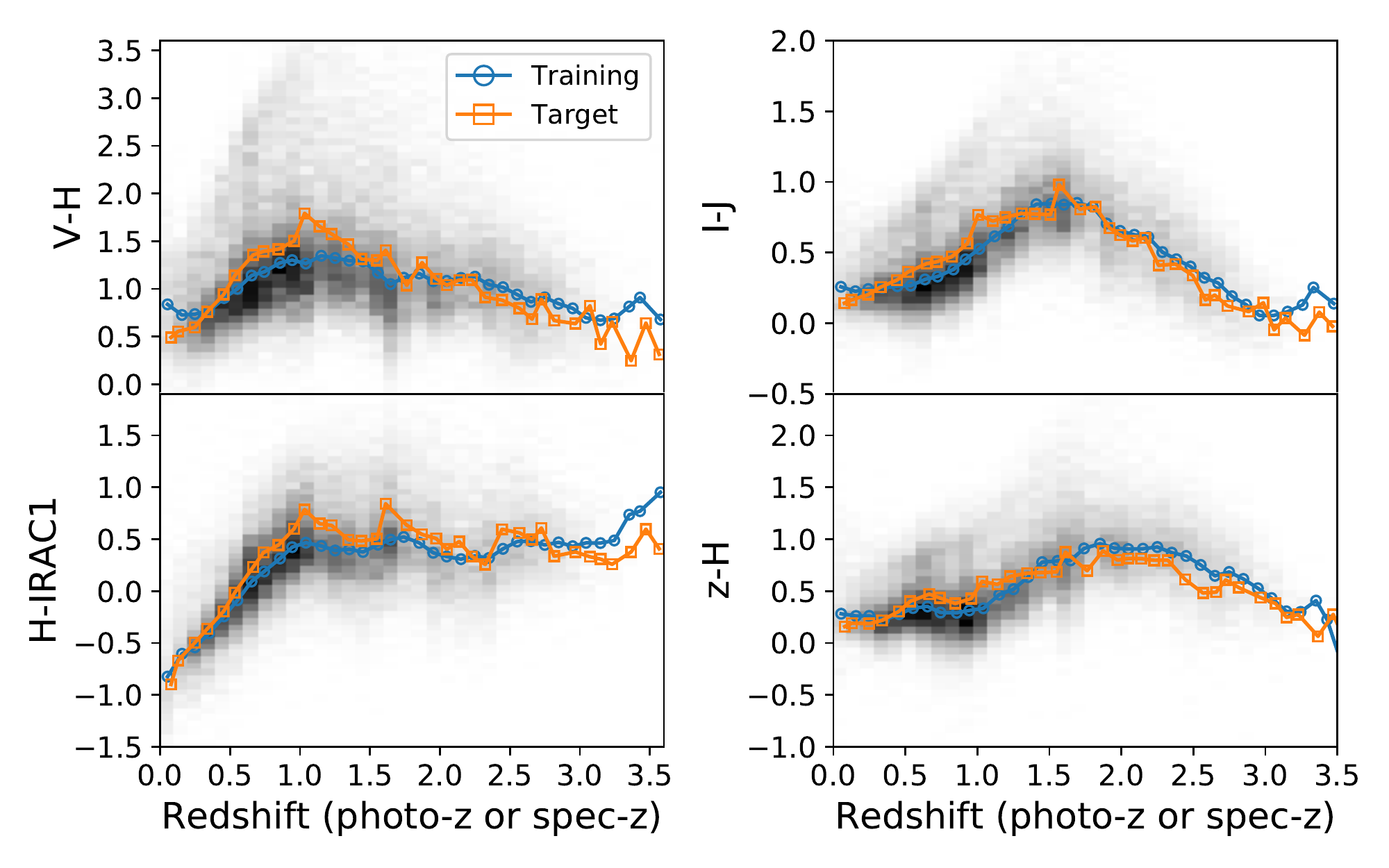}\vspace{-0.3cm}
\caption{Observed colors in AB mag, $V-H$, $I-J$, $z-H$, $H-$IRAC1 vs redshift of the training and target sample. The greyscale is the number density of galaxies, i.e. the darker color represents a denser region. Blue and orange points represent the median colors at each redshift bin ($\Delta z = 0.1$) for the training and target sample, respectively. The distribution of observed colors of the target sample is similar to that of the training sample. There is a clear correlation between color and redshift with $V-H$, $I-J$, and $z-H$ peaking at about $z\sim1.1$, $1.6$, and $1.9$, respectively, when the Balmer break falls between the two bands. H-IRAC evolves strongly at $z < 1$ due to the $1.6 \mu m$ bump. }
\label{fig:zcolor}
\end{figure*}

Figure~\ref{fig:zcolor} illustrates four observed colors, $V-H$, $I-J$, $H-$IRAC1, and $z-H$, as a function of galaxy redshift (grey points), shown along with the median colors for the training (blue) and target (orange) samples. IRAC1 corresponds to the {\it Spitzer}/IRAC 3.6\,$\,\mu$m band.
Most of the galaxies in the target sample have similar colors as the training sample, except one galaxy outside the $V-H$ range and three galaxies out of $H-$IRAC1 range.
Clearly, there is a smooth and continuous redshift evolution through color space, showing how multicolor measurements can constrain redshift estimates. Each color varies with redshift in a different way, reaching a maximum at different redshifts. For example, the peak of the $I-J$ color is about $z=1.6$ when the Balmer break ($4000\textrm{\AA}$) falls between the I and J bands. For the same reason, $V-H$ peaks at $z=1.1$, and $z-H$ peaks at $z=1.9$. $H-$IRAC1 also shows
a strong evolution with redshift, especially over $z < 1$, due to the $1.6 \mu m$ bump in galaxy SEDs arising from H$^{-}$ opacity \citep{saw02}. However, single colors are degenerate in redshift as one can see from the scatter in the greyscales. Stellar population age and dust extinction are degenerate in most optical/NIR color space such that dusty starburst galaxies can be difficult to distinguish from old, passive galaxies. To constrain redshifts, there only needs to be a distinct spectral feature in the SED of the galaxy. For most such galaxies, it tends to be the 1.6\,$\mu$m bump, so we do not think the age-extinction degeneracy introduces a significant limitation to our redshift estimation. We therefore use three color information to minimize degeneracies and use the multi-wavelength information to apply priors on galaxy SEDs.

\subsection{Cubes in the three dimensional color space}

In this study, we settled on using the following three colors after experimenting with various combinations of color space: 
\begin{align*}
~~(V-H) & = [ -1.53,~5.07],~(I-J) = [-2.1,~3.5] \\
 \& & ~ (z-H)  =  [-3.1,~6.9] 
\end{align*}
where, the color ranges are defined based on CANDELS photometry of the training sample.

Although the relationship between color and redshift associated with photometric signatures is crucial, we find that the choice of colors strongly relies on the quality of photometry as well. For instance, even though $H-$IRAC1 is useful to constrain redshifts at $z<1$, we find that $z-H$ improves photo-z estimates better than $H-$IRAC1 because the 3.6\,$\mu$m imaging is shallower than the $z$ band in our dataset. In the template
models, the band is also contaminated by 3.3 $\mu m$ emission from polycyclic aromatic hydrocarbons and hot dust continuum whose intensity relative to the stellar population is rather poorly calibrated. Furthermore, the 3.6\,$\mu$m bandpass will not be available over the wide fields of {\Euclid} and {\it Roman}.  $U-B$ is also important for a secure detection of the Balmer break at low redshifts and for detecting the Lyman break at $z>2$. However, $U$ and $B-$bands in CANDELS are from ground-based telescopes (except F435W in the GOODS-S and -N fields), have shallower depth and lower resolution, so give weaker constraints on photo-z estimates. We also tried to classify sources using galaxy morphological parameters such as $Gini$, $M_{20}$, and half-light radius, and a H-band magnitude. However, any combinations of morphologies and/or a magnitude, with colors, yielded worse results in deriving photometric redshifts.

\begin{figure*}
\centering
\includegraphics[width=6in]{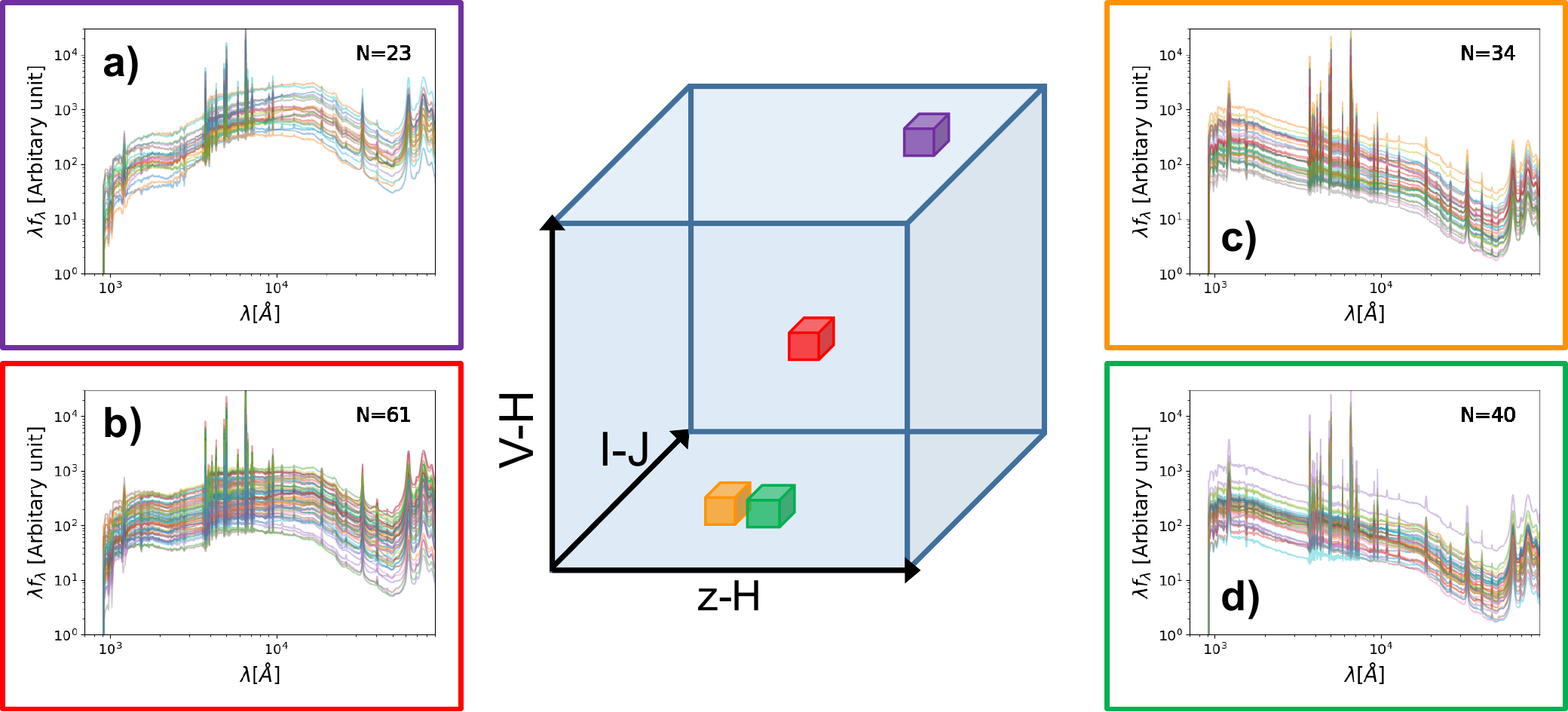}
\caption{An illustration of the color-cube-based template library made with the training sample in three-dimensional color space. In the middle, there are four $0.2\times 0.2 \times 0.2$ mag size cubes at different locations in the $V-H$, $I-J$, and $z-H$ color space. 
We also show the range of galaxy templates within the four color cubes with the number of rest-frame SEDs in the cube; a) for the purple cube, b) for the red, c) for the orange, and d) for the green cube. For example, 23 galaxies in the training sample are assigned to the purple cube so that the purple cube-based template (a) consists of 23 rest-frame SEDs (redshifted best-fit SEDs) of those galaxies. The red, green, and purple cubes have different types of galaxy SEDs. However, since the orange cube is located next to the green cube, their color-cube-based templates are very similar. If a target galaxy is located in the red cube, its photometric redshift is determined using the galaxy templates (61 galaxy SEDs in b) in the red box.} 
\label{fig:cube}
\end{figure*}

Using $V-H$, $I-J$, and $z-H$ colors, we simply make bins in each color with a bin size of $\Delta = 0.2$ mag to build $0.2\times0.2\times0.2$ size cubes in three dimensional color space as shown in the middle of Figure~\ref{fig:cube}.
As a result, all of the galaxies in the training sample are assigned to 1,286 cubes in $V-H$, $I-J$, and $z-H$ color space. 
The bin size is chosen because it is well matched to the photometric scatter of a typical $\sim$5$\sigma$ source. 
In addition, using a larger bin size with less cubes results in heterogeneous templates within a cube. If on the other hand, 
one adopts more cubes with a smaller bin size, many cubes are not populated with templates from the training sample and
the library of templates within each cube is sparsely sampled. 

\subsection{Best-fit SEDs of the training sample}

We next need to obtain the best-fit spectral energy distribution (SED) of each galaxy in the training sample. This would then develop the library of templates applicable to each color cube in three-color space. 

In order to obtain the best-fit SEDs for the training set of galaxies, we fix the redshift of each galaxy to its CANDELS median photo-z and fit model templates to its multi-wavelength photometry. It is well-known that adding zero-point offsets produce both lower scatter and outlier fraction of photometric redshifts \citep{dah13}. We, therefore, apply the zero point offsets in Table~\ref{table:zp} to the observed photometry of the training sample before doing the fits. The origin of this "zero point offset" is debatable since the instruments are calibrated with greater precision than the values estimated. Physically, it may be because of ignorance of galaxy templates at that level, particularly nebular 
emission. However, it could just as well be because of differences in how photometry is done; for instance, \cite{ske14} found that on the same data, two different cataloging techniques find differences at the $0.1-0.3$ mag level. Distinguishing between these scenarios is beyond the scope of this work. In Appendix A, we compute the zero point offsets for each individual filter and each field on the sample with spec-zs. 

For the fitting, we use the EAZY \citep{bra08} code that has been widely used for large surveys, such as COSMOS \citep{muz13}, CANDELS \citep{dah13}, and 3D-HST \citep{ske14}. EAZY has 12 galaxy templates derived from the Flexible Stellar Population Synthesis (FSPS) models \citep{con09, con10} trained on the UltraVISTA photometric catalogs \citep{muz13} with the method used for the original EAZY templates. Thus, we use the linear combination of those 12 galaxy templates derived from FSPS. We also apply the default rest-frame template error function scaled by a factor of 0.2, which helps to account for systematic wavelength dependent uncertainties in the templates (see the details from the EAZY manual\footnote{http://www.astro.yale.edu/eazy/eazy\_manual.pdf}). 

Among  39,391 best-fit SEDs of the training sample, we exclude the 4.6\% of fits having bad $\chi^{2}$ to avoid using poor SED fits to the data. We then shift the best-fit SEDs to rest-frame wavelengths with CANDELS median photo-zs (hereafter, rest-frame SEDs).

\begin{table*}
\centering
\caption{Photometric redshift results with LSST+\Euclid-deep/{\it Roman}-quality photometry ($ugrizYJH$ bands).\label{tab:gs8}}
\begin{tabular}{|c|c|c|c|c|c|c|c|c|c|}
\hline \hline
\multirow{2}{*}{Field Name}  & \multirow{2}{*}{$z_{spec}$ range} & \multirow{2}{*}{N} & \multicolumn{3}{c|}{THIS WORK} && \multicolumn{3}{c|}{EAZY} \\ \cline{4-10} 
 &  &  & $\sigma_{full}$ & $\sigma_{NMAD}$ & $f_{out}$ && $\sigma_{full}$ & $\sigma_{NMAD}$ & $f_{out}$ \\ \hline
\multirow{3}{*}{GS} &  all & 1655 & 0.194 & 0.026 & 0.063 && 0.194 & 0.029 & 0.073 \\ 
                                   & $z < 2$  & 1410 & 0.170 & 0.025 & 0.038 && 0.164 &0.026 & 0.045 \\ 
                                    & $ z > 2$ & 245 & 0.269 & 0.037 &  0.208 && 0.287 & 0.047 & 0.233 \\ \hline
\multirow{3}{*}{GN} & all & 1343 & 0.157 & 0.028 & 0.056 && 0.161 & 0.028 & 0.056 \\ 
                                 & $z < 2$ & 1129 & 0.131 & 0.027 & 0.032 && 0.132 & 0.026 & 0.029 \\
                                   &$z > 2$ &  214 & 0.234 & 0.035 & 0.187 && 0.247 & 0.044 & 0.200 \\ \hline
\end{tabular}
\vspace{-0.3cm} 
\end{table*}

\subsection{Color-cube-based template library}

We construct the color-cube-based template library using the rest-frame SEDs of the training sample. The rest-frame SED of each galaxy in the training sample is assigned to its designated cube in $V-H$, $I-J$, and $z-H$ color space as illustrated in Figure~\ref{fig:cube}. The color-cube-based template library is basically a collection of galaxy SEDs with similar three-color properties. In Figure~\ref{fig:cube}, as an example, we show four sets of galaxy templates of four cubes located differently in $V-H$, $I-J$, and $z-H$ space. Purple, blue, and green boxes have different types of galaxy SEDs representing mostly blue star-forming galaxies (green) or red passive galaxies (purple). However, if the cubes are located closely, their SEDs are very similar like green and orange cubes. 
We note that there are a maximum of 1109 rest-frame SEDs in a cube. 

By constraining the range of templates applicable to a galaxy using three-color space with a photo-z training sample, we are constraining the range of extinction values that may be possible among high redshift galaxies, and constraining the range of star-formation histories in each redshift bin, which helps alleviate color-redshift degeneracies.

\subsection{Photometric redshift estimates}

There are two main parts in the algorithm for determining the photometric redshift of a target sample galaxy using the color-cube-based template library. The first is to find the correct color cube of the target galaxy in three dimensional color space. The second is to estimate the photo-z of the target galaxy by fitting its multiwavelength photometry with the range of color-cube-based templates with redshift as a free parameter.
This is determined by minimizing the $\chi^{2}$ between the observed SED and the scaled color-cube-based templates within that cube.

The photo-z here is determined using the EAZY code modified to use the color-cube-based template library. As before, the photometry of the target galaxy is corrected by the zero-point offsets (Table~\ref{table:zp}). We do not allow linear combination of the library of galaxy templates as is done by EAZY. For example, if three galaxies are located in the purple cube in Figure~\ref{fig:cube}, our method determines photo-zs of them only using the 23 galaxy templates in the purple box (plot a) and not a linear combination of subsets of the 23 templates. In other words, the rest-frame SEDs of the training sample of galaxies are assumed to be representative of, and applicable to the target sample of galaxies; the photo-z of the target is thereby derived from the best matching template. The photo-z of the target galaxy is the peak of the redshift probability distribution function (i.e. $z_{peak}$ in EAZY terminology).

One can apply the same color ranges/combinations and the training sample to any survey. While it is possible that there could be extremely red or blue sources that are different from the training sample with CANDELS, those sources would be very rare that it would not affect the photo-z precision statistics and one could use the nearest matching color cube to provide a template library.

\section{Photo-z performance and validation}

We test the quality of photometric redshifts computed using the color-cube-based template library. We compare our photo-z estimates of the target sample of galaxies with their spectroscopic redshifts at $0 < z < 6$ and to the photometric redshifts derived using EAZY on the same sample. Although CANDELS already has photo-z estimates using EAZY with the original templates from \cite{bra08}, we re-derive photo-zs using EAZY's 12 FSPS galaxy templates with the same settings discussed in Section 3.2. We find that EAZY results in very similar or slightly better photo-z estimations compared to median CANDELS photo-z. 
Since the only difference between EAZY and our analysis is that we use a more constrained template library based on our training sample, we can validate the performance of our cube-based template library by comparing to the results from EAZY. 

In this section, we investigate several implementations of our color-cube-based template library using only 8 bands (LSST+Euclid-deep/{\it Roman}-like) and an inclusion of longer wavelengths, Ks (2.2 $\mu m$) or IRAC 1 (3.6 $\mu m$).

\begin{figure*}
\centering
\includegraphics[width=6in]{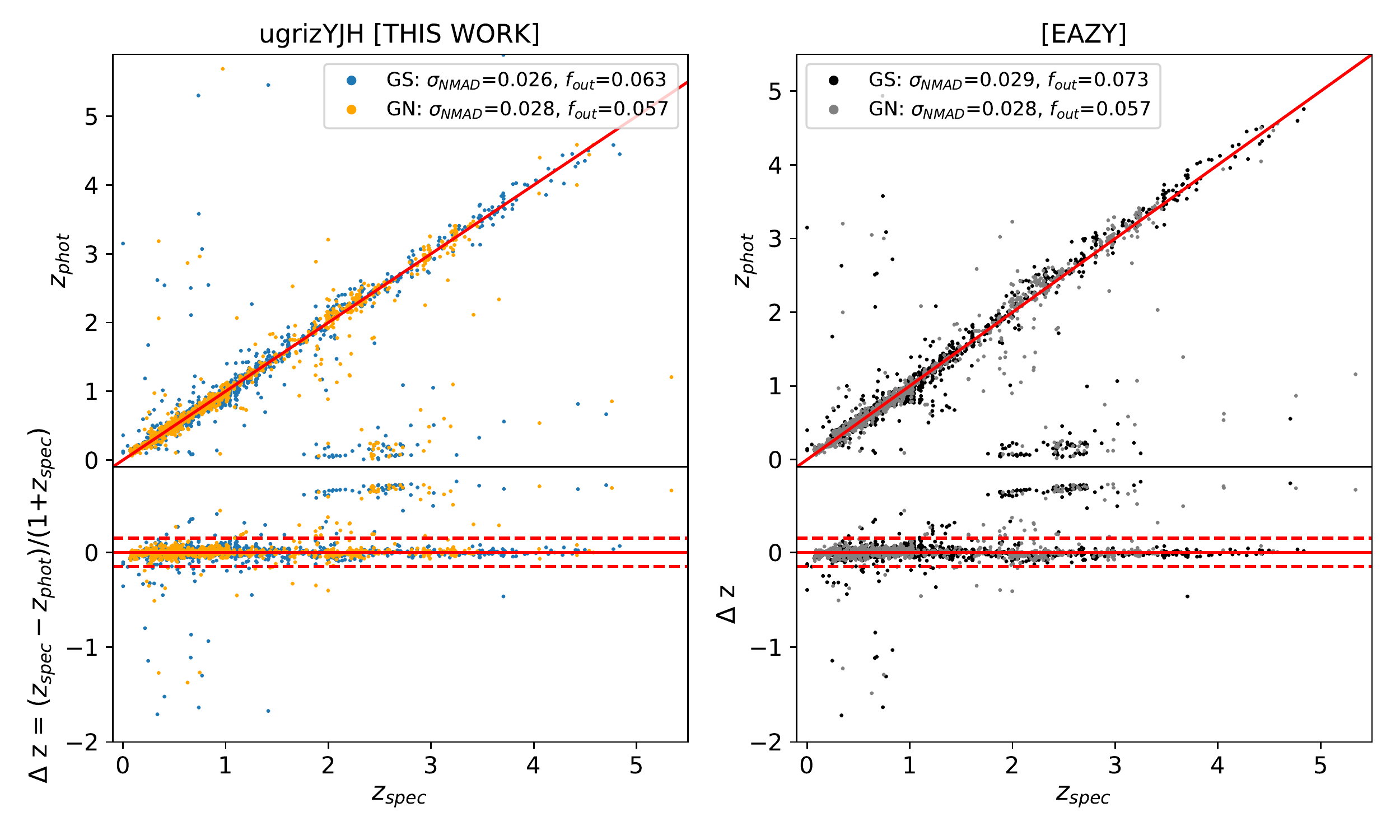}
\caption{Comparison between spectroscopic redshifts ($z_{spec}$) and photo-z estimates ($z_{phot}$) of the target sample in GOODS-S (GS) and GOODS-N (GN). Left: A result from our method with GOODS-S (blue) and GOODS-N (orange) galaxies. Right: The result from EAZY for the same galaxies (GS: black, GN: grey). We plot $z_{phot}$ vs $z_{spec}$ in the top panel and $\Delta_{z}$ ($=(z_{spec}-z_{phot}$)/(1+$z_{spec}$)) vs. $z_{spec}$ in the bottom panel with $\sigma_{NMAD}$ and the outlier fraction $f_{out}$ measured for each field. $\Delta_{z} = 0$ is shown as a red line with dashed lines indicating $|\Delta_{z}| = 0.15$. Galaxies located below/above the dashed lines are classified as outliers. Here, we use LSST+\Euclid-deep/{\it Roman}-quality multi-band photometry to derive photo-z using the color-cube-based template library made from $V-H$, $I-J$, and $z-H$ colors. Our method results in an improved photo-z precision of $\sigma_{NMAD} = 0.26-0.28$ and smaller outlier fraction of about 6\% comparing to EAZY with $\sigma_{NMAD} = 0.028-0.029$ and $f_{out} =$ 6-7\%.} 
\label{fig:pz8}
\end{figure*}


\subsection{Results with ugrizYJH photometry}

The results of our photo-z fits are shown in Table~\ref{tab:gs8} with statistics quantifying the accuracy of the photo-z estimates. Root mean square (RMS) of $\Delta_{z}$, normalized median absolute deviation (NMAD) of $\Delta_{z}$, and outlier fraction ($f_{out}$) are defined in Equation 1. 
\begin{align}
\Delta_{z} &= (z_{spec}-z_{phot})/(1+z_{spec}) \\
\sigma_{full} &=  {\rm rms}[\Delta_{z}] \nonumber \\
\sigma_{NMAD} &= 1.48 \times {\rm median}[|\Delta_{z}|]  \nonumber \\
f_{out} & =  N [ |\Delta_{z}| > 0.15] / N_{obj} \nonumber
\end{align}
where, the outlier fraction, $f_{out}$ is the fraction of outliers defined as objects with $|\Delta_{z}| > 0.15$. Figure~\ref{fig:pz8} compares our photo-z estimates to spec-zs of galaxies in the two fields with $\sigma_{NMAD}$ and $f_{out}$ listed for each field. Note that the number of target galaxies are reduced to 1,655 and 1,343 in GOODS-S (GS) and -N (GN), respectively, since we are not able to obtain photo-zs of 2.4\% galaxies in GS and 0.9\% galaxies in GN because of two reasons. 
One, some of the target sample falls in cubes which do not have good templates because those cubes are only populated by SEDs with bad fits in the training sample. Two, we find that 1.6\% of the target sample are not located in the training cubes; about half of these are at $z_{spec} > 3.6$ and comprise about 32\% of the $z_{spec} > 3.6$ target galaxies.

The color-cube-based template library is applicable to all redshift ranges although the signal-to-noise ratio requirement in U-band 
limits the training sample to $z < 3.6$. We find that only three $z_{spec} > 3.6$ galaxies in the target sample are excluded because they fall outside the color range of the training sample with S/N $> 3$ in U-band. If we try to derive redshifts of galaxies not located in the training cubes using the color-cube-based template library of the nearest populated color cube, we find that it results in worse estimates. However, the CANDELS spec-z sample has a surprising lack of high redshift galaxies with only 115 galaxies ($\sim 2\%$ of the spec-z sample) at $z_{spec} > 3.6$ and 74\% of them are in the GOODS-S and -N fields. This is rather inadequate to investigate the photo-z performance in this redshift range and we therefore do not present the results for galaxies not located in the training cubes. In the future, we will augment the target and training samples with the results of current spectroscopic surveys, particularly at $z_{spec} > 3$.

In Table~\ref{tab:gs8}, we compare $\sigma_{full}$, $\sigma_{NMAD}$, and $f_{out}$ with EAZY results. We find that our method results in smaller $\sigma_{full}$, $\sigma_{NMAD}$, and $f_{out}$ compared to EAZY, indicating the improvement of photo-z estimates.  At $0 < z < 6$, $\sigma_{NMAD}$ of this study is 0.026 (0.028), while EAZY yields 0.029 (0.028) in GS (GN). In Figure~\ref{fig:pz8}, outliers residing below/above the red dashed lines correspond to $|\Delta_{z}| > 0.15$. We find that  $f_{out}$ is about 6\%, similar to EAZY. $\sigma_{full}$ is 0.194 and 0.157 in GS and GN. We improve the photo-z precision by about 10\%  and the outlier fraction is decreased by about 13\% compared to EAZY in GS, while in GN $f_{out}$ and $\sigma_{NMAD}$ are almost identical between the two techniques. However, we find that at $z > 2$, our method shows a significant improvement on the photo-z estimates in both GS and GN as shown in Table~\ref{tab:gs8}. At $z <2$,  our method results in smaller statistics compared to EAZY with galaxies in GS, while $\sigma_{NMAD}$ and $f_{out}$ in GN are slightly larger than ones from EAZY. Overall, we obtain better photo-z performance with GS compared to GN. This might be due to the fact that U-band in GS is slightly deeper than GN. The 
5 $\sigma$ limiting depth of U-band VLT/VIMOS in GS is 27.97 AB mag while GN U-band KPNO/Mosaic has 5 $\sigma$ of 26.7 AB mag. With the color-constrained templates limiting the range of extinctions, the deeper U-band of GS can classify high and low redshift galaxies better in the fitting.

\begin{figure}
\centering
\includegraphics[width=0.85\columnwidth]{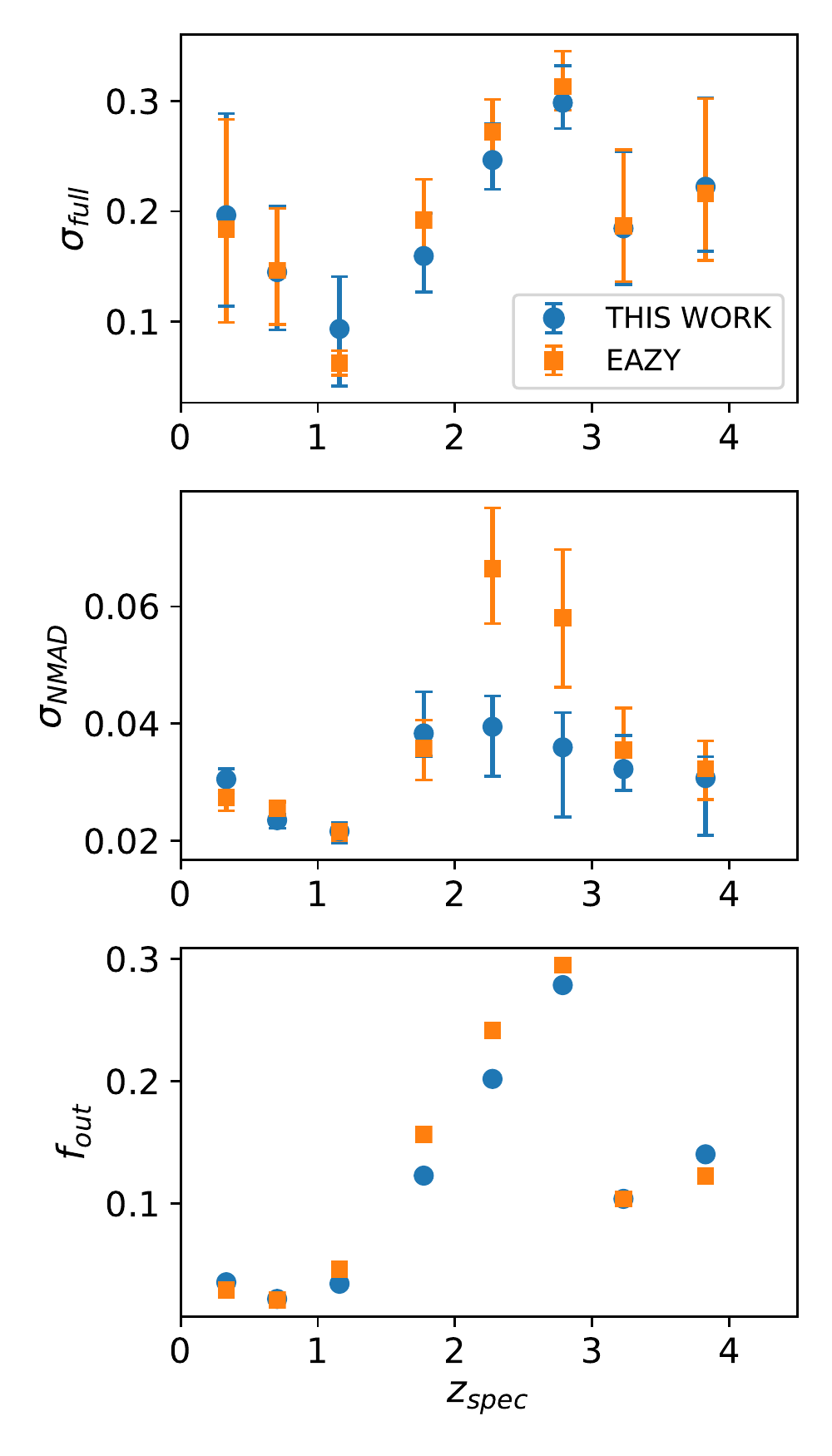}\vspace{-0.3cm}
\caption{Three photo-z statistics, $\sigma_{full}$, $\sigma_{NMAD}$, and $f_{out}$ as a function of spec-z ($z_{spec}$). Bin size is $\Delta z_{spec} = 0.5$, while the last bin include galaxies at $3.5 < z_{spec} < 5.5$. Statistics measured in each redshift bin are shown on the Y-axis for this work (blue) and EAZY (orange). X-axis is the median $z_{spec}$ in each bin. The number of galaxies in each redshift bin is 644, 1223, 493, 179, 203, 122, 77, 57, respectively. Error bars of $\sigma_{full}$ and $\sigma_{NMAD}$ are determined by bootstrapping the sample. Our photo-z performance at $1.5 < z < 3.5$ is better with smaller values of $\sigma_{full}$, $\sigma_{NMAD}$, and $f_{out}$ than ones from EAZY, while similar to EAZY at $z < 1.5$ and $z > 3.5$. In particular, our method has a very high precision with $\sigma_{NMAD} <0.04$ at $z>2$ compared to EAZY with $ < 0.06$, indicating about $30\%$ improved performance at $z>2$.}
\label{fig:pz8_zbin}
\end{figure}

We also plot $\sigma_{full}$, $\sigma_{NMAD}$, and $f_{out}$ as a function of redshift in Figure~\ref{fig:pz8_zbin}. At all redshifts considered here, photo-z precision is relatively high with $0.02 < \sigma_{NMAD} < 0.04$. In particular, at $z>2$, $\sigma_{NMAD}$ of our method is significantly small, $\sigma_{NMAD} \sim 0.03-0.04 $, comparing to EAZY with $\sim 0.06$ at $2< z <3$, indicating that we can improve the photo-z precision by about 30\% by using the color-cube-based template library at $z>2$. Other statistics, $\sigma_{full}$ and $f_{out}$ are smaller than ones from EAZY at $1.5 < z < 3.5$. They are however, similar to EAZY at lower redshifts and the highest redshift bin ($z>3.5$). Our result shows that the color-cube-based template library is powerful for estimating photometric redshifts when the number of photometric bands is limited, especially at high redshifts. 

\subsection{Results with an inclusion of IRAC 1 or Ks bands}

It has been shown that the derivation of photometric redshifts benefit from having multiple narrow bands and having broad wavelength coverage \citep{dah10}, especially if they straddle strong features in galaxy SEDs such as the Balmer break and 1.6$\mu m$ bump. It is therefore worth assessing if the addition of longer wavelength bands
to the planned ugrizYJH photometry would be beneficial for photo-z derivation. We investigate here the redshift accuracy when we add longer near-infrared bands, Ks (2.2 $\mu m$) or IRAC 1 (3.6 $\mu m$) to the $ugrizYJH$ bands. 

We first derive photo-zs using our method by including the IRAC 1 band, i.e. 9 bands adding IRAC 1 to the $ugrizYJH$ bands. The results are shown in Table~\ref{tab:gs9}. In comparison to the result only using ugrizYJH (Table~\ref{tab:gs8}), we are able to improve the overall performance of photo-zs when the IRAC 1 band is included. We obtain a smaller $\sigma_{NMAD}=0.025$ in both GS and GN. The fraction of outliers decreases from 6.3\% to 5.1\% for the GS and 5.6\% to 5.2\% for the GN. Likewise, we find $\sigma_{full}$ drops dramatically in both fields, to 0.141 (GS) and 0.130 (GN), which corresponds to about a 27\% and 17\% decrease, respectively. This implies that the inclusion of the IRAC 1 band has a significant impact on reducing outliers. This is particularly relevant for the 40\,deg$^{2}$ {\it Euclid}-deep fields which are the target of {\it Spitzer} observations achieving depths of 5$\sigma$=24.6 AB mag (PI: P. Capak), although this depth is shallower than the IRAC data used here.

Next, we derive photo-zs with 9 bands adding Ks band to the ugrizYJH. Ks band photometry in GS is quite deep with VLT/HAWK-I (5 $\sigma$ depth of $26.45$ AB), while the GN Ks-band is shallower, $\sim 24.4$ AB mag. As shown in Table~\ref{tab:gs9}, the inclusion of the shallow Ks band photometry has only a marginal effect for galaxies in GN when compared to the result without Ks-band. The relatively low impact on the photometric redshifts when including Ks-band could be attributed to the fact that the H band from \HST\ already provides a tight constraint on the SED longwards of the Balmer break and the 1.6$\mu m$ bump has a relatively small effect on photo-z estimation at the redshift range of interest. Despite having deeper Ks band in GS, we obtain worse results with a photo-z precision of $\sigma_{NMAD}=0.29$. The origin of this discrepancy is a puzzle and might require a careful re-evaluation of the HAWK-I photometry. 

\begin{table}
\centering
\caption{Derived Photo-z precision when including IRAC1 or Ks bands\label{tab:gs9}}
\begin{tabular}{ccccc}
\hline \hline
 & Field name &  $\sigma_{full}$ & $\sigma_{NMAD}$ & $f_{out}$ \\ \hline
\multirow{2}{*}{with IRAC 1} & GS &  0.141 & 0.025 & 0.051 \\ 
 & GN & 0.130 & 0.025 & 0.052 \\ \hline
\multirow{2}{*}{with K band} & GS &  0.168 & 0.029 & 0.063 \\  
 & GN &  0.158 & 0.027 & 0.060 \\ \hline
\end{tabular}
\vspace{-0.2cm}
\end{table}

\section{Conclusion: an introduction of new template libraries }

We investigate a new method for improving the accuracy of photo-z estimates for future wide area surveys which will be deep and have limited multi-wavelength coverage. There are two unique features in our method. 
To have a large unbiased galaxy sample, we select the training sample of galaxies from CANDELS to have photometric redshifts which are a median of 11 different photo-z codes. Although the survey area is small (about 0.25$deg^2$)  compared to future wide-area surveys, CANDELS is the only survey having de-blended photometry including deep near-IR bands and having spatial resolution comparable to {\it Roman} and $\Euclid$.
We then populate three-dimensional color space, $V-H$, $I-J$, and $z-H$ with the best fit SEDs for this training sample to derive a color-cube-based template library trained with high-quality multi-wavelength CANDELS photometry. 

We apply our color-cube-based template library to a target sample of galaxies with ugrizYJH photometry (LSST+ \Euclid/{\it Roman}-like) and derive their photo-zs. We quantify the performance of our method for determining photo-zs by comparing with spec-zs. We derive a photo-z precision of $\sigma_{NMAD}= 0.026-0.028$ and outlier fraction of about 6\% for galaxies in the GOODS-S and -N fields. We compare our results with EAZY using their templates and find that our method improves the photo-z precision by about 10\% and reduces the outliers by about 13\%. 
In particular, the color-cube-based template library is very effective at $2 < z < 3$ with significantly small $\sigma_{NMAD}$ 
of $\sim 0.04 $, compared to EAZY which yields $\sigma_{NMAD} \sim 0.06$. We also investigate the inclusion of additional longer wavelength photometry, e.g. Ks-band or IRAC1 (3.6\,$\mu$m), to the ugrizYJH photometry. Including deep IRAC1 photometry results in smaller $\sigma_{NMAD}=0.025$ and has the most significant impact on reducing outliers. 
However, the inclusion of Ks-band has marginal to no effect on the photometric redshifts mainly due to the fact that the deeper H-band data from {\it HST} already provides a tight constraint on the galaxy SEDs.

The main advantage of having a photo-z training sample is that it allows the SEDs of dusty galaxies to be constrained and distinguished from high-z Lyman break galaxies. Having purely a spec-z training sample would make it challenging to constrain the SEDs of such galaxies since spec-z samples tend be biased against dusty galaxies. This allows for better redshift determination of low-z dusty galaxies which would be
otherwise erroneously fitted to be high-z galaxies. Furthermore, having such a photo-z training sample constrains the range of extinction values that may be possible among high-z galaxies and constrains the range of star-formation histories in each redshift bin. Both these effects cause shifts in the best estimate photometric redshifts which is reducing the scatter at high-z. While it is possible that there could be sources with SEDs that are radically different from the training sample as one expands to wider areas, those sources will likely fail but would be rare enough that it would not affect the photo-z statistics.

Overall, our result clearly shows that the color-cube-based template library trained with CANDELS data in multi-dimensional color space is a powerful tool for determining photometric redshifts using the template-fitting approach, particularly for future large extragalactic surveys which will provide deep ugrizYJH photometry. 
These results will be highly applicable to the LSST and {\it Euclid}-deep fields which span 40 deg$^{2}$ and span similar depths as those sampled in this study.
They can also be applied to the LSST+{\it Roman} high latitude survey sky area which will span 2200 deg$^{2}$. 

\section*{Acknowledgements}
This work is funded by NASA/{\it Euclid} grant 1484822 and Joint Survey Processing (JSP) effort at Infrared Processing and Analysis Center (IPAC), which has been provided by NASA grant 80NM0018F0803. The data used in this work is based on observations taken by the CANDELS Multi-Cycle Treasury Program with the NASA/ESA \HST, which is operated by the Association of Universities for Research in Astronomy, Inc., under NASA contract NAS5-26555. The authors thank Andreas Faisst for thoughtful comments which improved this manuscript and Gabe Brammer for helpful ideas that improved our analysis. We also thank the anonymous referee for very useful comments that helped to improve the presentation of the paper.

\section*{Data availability}
The CANDELS multi-wavelength catalogs and photometric redshifts catalogs used in this study are publicly available on the Mikulski Archive for Space Telescopes
(MAST): https://archive.stsci.edu/prepds/candels/.
The photometric redshift estimations underlying this article will be shared on reasonable request to the corresponding author.

\appendix

\section{Zero point offsets}

As described in \cite{ilb06}, \cite{dah13}, and \cite{ske14} , the application of offsets to the photometry can improve photometric redshift fits. These are called zero point offsets but in general,
since the instrument photometry is calibrated with high accuracy, is unlikely to be due to actual errors in zero points. It is therefore
thought that any offsets are likely due to the mismatch between the SED templates and real galaxy SEDs \citep{guo13} although inconsistent isophotal apertures across bands may also be a contributing factor.
Zero point offsets are determined for individual filters and vary in different fields. We measure the zero point offset of each band by fitting the EAZY FSPS templates to the observed photometry of galaxies with spectroscopic redshifts as in \cite{ske14}. The offsets are computed iteratively by minimizing the median differences between the measured photometry and best fit templates integrated over the bandpasses in a particular field. This correction is then applied to the photometry of all sources. In principle, the correction should be calculated on a spec-z sample which is distinct from the sample on which the quality of photo-z is measured. However, due to the limited number of spec-zs available, these corrections are applied on the same sample. With future spectroscopic surveys, this is likely to be improved.

The listed zero point offset corrections in Table~\ref{table:zp} have been applied to the photometry when we obtain the best-fitted SED of each galaxy in the training sample in Section 3.2 as well as when we derive the photometric redshifts of the target sample using the color-cube-based template library in Section 3.4. The values in the table are multiplicative numbers to be applied to the flux densities in the corresponding filters.

\setcounter{table}{0}
\renewcommand{\thetable}{\Alph{section}\arabic{table}}
\begin{table*}
\caption{Zero point offsets for individual filters in 5 fields \label{table:zp}}
\begin{center}
\begin{tabular}{cccccccccc}
\hline \hline
\multicolumn{2}{c}{GOODS-S} & \multicolumn{2}{c}{GOODS-N} & \multicolumn{2}{c}{UDS} & \multicolumn{2}{c}{EGS} & \multicolumn{2}{c}{COSMOS} \\[0.1cm] 
\cline{1-2} \cline{3-4} \cline{5-6} \cline{7-8} \cline{9-10} 
& & & & & & & & & \\[-0.2cm]
Filter& zp offset & Filter & zp offset & Filter & zp offset & Filter & zp offset & Filter & zp offset \\
\hline
U Blanco & 1.0404 & U KPNO & 0.9483 & U CFHT & 1.1115 & u CFHT & 0.9980  & u CFHT & 0.9070  \\
U VLT & 1.0304 & F435W & 1.0616 & B Subaru & 0.9999 & g & 0.9619 & g & 0.9114 \\ 
F435W & 0.9855 & F606W & 0.9928 & V & 0.9902 & r & 0.9960 &  r &  0.9506\\
F606W & 0.9929 & F775W & 0.9811 & R & 0.9144 &  i & 0.9786 & i & 0.9415 \\
F775W & 0.9892 & F814W & 0.9826 & i &  0.9878 &  z & 0.9763 & z &  0.9503\\
F814W & 0.9910 & F850LP & 0.9531 & z & 0.9960 & F606W & 0.9315 &  B Subaru & 0.9888 \\ 
F850LP & 0.9674 & F105W & 0.9900 & F606W & 1.0518 & F814W & 0.9624 & F606W & 0.9476 \\ 
F098M & 1.0210 & F125W & 1.0020 & F814W & 0.9834 & F125W & 1.0025 &  F814W & 0.9784 \\
F105W & 1.0053 & F140W & 0.9915 & Y HAWK-I & 1.0144 & F140W & 0.9907 &  F125W &  1.0183\\
F125W & 1.0070 & F160W & 1.0000 & F125W & 1.0065 & F160W & 1.0000 & F160W  & 1.0000 \\ 
F160W & 1.0000 & Ks Subaru & 0.9540 & F160W & 1.0000 & J1 NEWFIRM & 0.9255 &  Y UVISTA & 0.9012 \\ 
Ks VLT/ISAAC & 1.0675 & Ks CFHT & 0.9373 & Ks VLT/HAWK-I& 1.0016 & J2 & 0.9117 & Ks UVIST & 0.9202\\
Ks VLT/HAWK-I & 1.0100 & IRAC 1 & 1.0070 & K UKIRT/WFCAM& 1.0039 & J3 & 0.9248& J1 NEWFIRM & 0.8942\\
IRAC 1 & 0.9616 & IRAC 2 & 0.9871 & IRAC 1& 1.0006 &  H1 & 0.9060 & J2  & 0.9034\\
IRAC 2 & 1.0175 & IRAC 3 & 1.0404 & IRAC 2 & 0.9825 &  H2 & 0.9316 & J3  & 0.9223\\
IRAC 3 & 0.9940 & IRAC 4 & 1.0507 & IRAC 3 & 0.9447 & Ks WIRCAM  & 0.9636 & H1 & 0.8969\\
IRAC 4 & 1.0116 &  &  & IRAC 4 & 0.9848 & IRAC 1  & 0.9857& H2 &  0.8777\\
 &  &  &  &  &  &  IRAC 2 & 0.9783& IRAC 1& 0.9372\\
 &  &  &  & &  &  IRAC 3 & 1.0594& IRAC 2& 0.9166 \\
 &  &  &  & & & IRAC 4 & 0.7614 & IRAC 3 & -- \\
  &  &  & & & &  & & IRAC 4 &  -- \\
 &  &  & & & &  & & IA484 Subaru &  0.9441 \\
&  &  & & & &  & & IA527  &  0.9582 \\
&  &  & & & &  & & IA624  &  0.9482\\
&  &  & & & &  & & IA679  &  0.7946 \\
&  &  & & & &  & & IA738 &  0.9942 \\
&  &  & & & &  & & IA767 &  0.9608 \\
&  &  & & & &  & & IA427 & 0.9452 \\
&  &  & & & &  & & IA464 &  0.9458\\
&  &  & & & &  & & IB505 &  0.9414\\
&  &  & & & &  & & IB574 &  0.9749\\
&  &  & & & &  & & IB709 &  0.9618 \\
&  &  & & & &  & & IB827 &  0.9809 \\ \hline
\end{tabular}
\end{center}
\end{table*}

\bsp	
\label{lastpage}
\end{document}